\documentclass[preprint2]{aastex}

\shorttitle{WASP-12b: the hottest transiting extra-solar planet yet discovered}
\shortauthors{Hebb et al.}

\begin{document}


\title{WASP-12b: the hottest transiting extra-solar planet yet discovered}

\author{L.~Hebb\altaffilmark{1},
A.~Collier-Cameron\altaffilmark{1},
B.~Loeillet\altaffilmark{2},
D.~Pollacco\altaffilmark{3},
G.~H\'ebrard\altaffilmark{4},
R.A.~Street\altaffilmark{5},
F.~Bouchy\altaffilmark{4,6},
H.C.~Stempels\altaffilmark{1},
C.~Moutou\altaffilmark{2},
E.~Simpson\altaffilmark{3},
S.~Udry\altaffilmark{13},
Y.C.~Joshi\altaffilmark{3},
R.G.~West\altaffilmark{7},
I.~Skillen\altaffilmark{8},
D.M.~Wilson\altaffilmark{9},
I.~McDonald\altaffilmark{9},
N.P.~Gibson\altaffilmark{3},
S.~Aigrain\altaffilmark{10},
D.R.~Anderson\altaffilmark{9},
C.R.~Benn\altaffilmark{8},
D.J.~Christian\altaffilmark{3},
B.~Enoch\altaffilmark{1},
C.A.~Haswell\altaffilmark{11},
C.~Hellier\altaffilmark{9},
K.~Horne\altaffilmark{1},
J.~Irwin\altaffilmark{12},
T.A.~Lister\altaffilmark{5},
P.~Maxted\altaffilmark{9},
M.~Mayor\altaffilmark{13},
A.J.~Norton\altaffilmark{11},
N.~Parley\altaffilmark{1},
F.~Pont\altaffilmark{10},
D.~Queloz\altaffilmark{13},
B.~Smalley\altaffilmark{9},
\and
P.J.~Wheatley\altaffilmark{14}
}

\altaffiltext{1}{School of Physics and Astronomy, University of St Andrews, North Haugh, St Andrews, Fife KY16 9SS, UK }
\altaffiltext{2}{Laboratoire d'Astrophysique de Marseille, BP 8, 13376 Marseille Cedex 12, France }
\altaffiltext{3}{Astrophysics Research Centre, School of Mathematics \&\ Physics, Queen's University, University Road, Belfast, BT7 1NN, UK}
\altaffiltext{4}{Institut d'Astrophysique de Paris, CNRS (UMR 7095) --  Universit\'e Pierre \&\ Marie Curie, 98$^{bis}$ bvd. Arago, 75014 Paris, France }
\altaffiltext{5}{Las Cumbres Observatory, 6740 Cortona Dr. Suite 102, Santa Barbara, CA 93117, USA }
\altaffiltext{6}{Observatoire de Haute-Provence, 04870 St Michel l'Observatoire, France }
\altaffiltext{7}{Department of Physics and Astronomy, University of Leicester, Leicester, LE1 7RH, UK }
\altaffiltext{8}{Isaac Newton Group of Telescopes, Apartado de Correos 321, E-38700 Santa Cruz de la Palma, Tenerife, Spain }
\altaffiltext{9}{Astrophysics Group, Keele University, Staffordshire, ST5 5BG, UK }
\altaffiltext{10}{School of Physics, University of Exeter, EX4 4QL, UK }
\altaffiltext{11}{Department of Physics and Astronomy, The Open University, Milton Keynes, MK7 6AA, UK}
\altaffiltext{12}{Harvard-Smithsonian Center for Astrophysics, 60 Garden Street, Cambridge, MA, 02138 USA}
\altaffiltext{13}{Observatoire de Gen\`eve, Universit\'e de Gen\`eve, 51 Ch. des Maillettes, 1290 Sauverny, Switzerland }
\altaffiltext{14}{Department of Physics, University of Warwick, Coventry CV4 7AL, UK}

\begin{abstract}
We report on the discovery of WASP-12b, a new transiting extrasolar planet
with $R_{\rm pl}=1.79 \pm 0.09 R_J$ and $M_{\rm pl}=1.41 \pm 0.1 M_J$.
The planet and host star properties were derived from a Monte Carlo Markov Chain
analysis of the transit photometry and radial velocity data.  
Furthermore, by comparing the stellar spectrum with theoretical spectra and stellar evolution models, 
we determined that the host star is a super-solar metallicity ([M/H]$=0.3^{+0.05}_{-0.15}$), 
late-F (T$_{\rm eff}=6300^{+200}_{-100}$~K) star which is evolving off the zero
age main sequence.  The planet has an equilibrium temperature of T$_{\rm eq}$=2516~K 
caused by its very short period orbit ($P=1.09$~days) around the hot, 12th magnitude host star.
WASP-12b has the largest radius of any transiting planet yet detected.   
It is also the most heavily irradiated and the shortest period planet in the literature.
\end{abstract}


\maketitle


\section{Introduction}

Transiting extra-solar planets have provided tremendous information about the properties 
of planets outside our Solar System.  Since 2006, a burst of new planet discoveries have
been reported.
We are now beginning to see the variety of exoplanets which exist in the Galaxy and to 
classify them based on their properties.  Furthermore, 
due to the increasing number of planets being discovered and due to the detailed, multi-wavelength
follow-up of a handful of very bright transiting systems (e.g.\ HD~209458, HD~189733), we are 
able to provide strong observational tests of theoretical models of planet formation and evolution.

Exotic planets are particularly important because they push the boundaries of 
our theoretical understanding.  HD~209458b, for example, confounded theorists
with its large radius \citep{brown01}.  Since its discovery,
a class of similar planets have been found suggesting these highly-irradiated,
low-density planets are not rare.  Here, we report on the discovery
of a new extreme transiting extra-solar planet with a short orbital
period, enlarged radius, and super-solar metallicity host star.

In this paper, we first describe all the observations we obtained to 
detect and analyse the transiting star-planet system (\S\ref{sec:observations}).  
We describe the data analysis in \S\ref{sec:analysis} 
where we present the properties of the planet and its host star.
Finally in \S\ref{sec:discuss}, we discuss the planet properties in the context of current
theoretical understanding of planet formation.

\section{Observations}
\label{sec:observations}

2MASS~J063032.79+294020.4 (hereafter WASP-12) is a bright F9V star.  It has been identified in
several northern sky catalogues which provide broad band optical \citep{NOMAD}
and infra-red 2MASS magnitudes \citep{2MASS} and proper motion information.  
Coordinates and broad band magnitudes of the star are given in Table~\ref{tab:specparms}.

\begin{table}
\caption[]{Stellar Parameters for WASP-12.  The broad band magnitudes are 
obtained from the NOMAD~1.0 catalogue. The stellar parameters are derived from 
our spectral synthesis of observed spectra of WASP-12 (see \S\ref{sec:specsynth}).}
\label{tab:specparms}
\begin{center}
\begin{tabular}{cc}
 Parameter    & WASP-12 \\
 \hline\\
${\rm RA (J2000)}$    &  06:30:32.79         \\
${\rm Dec (J2000)}$   & +29:40:20.4         \\
              &                      \\
${\rm B}$     &  $12.11\pm 0.08$     \\
${\rm V}$     &  $11.69\pm 0.08$     \\
${\rm I}$     &  $11.03\pm 0.08$     \\
${\rm J}$     &  $10.477\pm 0.021$   \\
${\rm H}$     &  $10.228\pm 0.022$   \\
${\rm K}$     &  $10.188\pm 0.020$   \\
              &                      \\
$T_{\rm eff}$ & $6300^{+200}_{-100}$ K     \\
$[$M/H$]$     &  $ 0.30^{+0.05}_{-0.15}$   \\
log\,$g$      &  $ 4.38  \pm 0.10$  	   \\
$v$\,sin\,$i$ & $<2.2 \pm 1.5$ km/s         \\
\hline\\
\end{tabular}
\end{center}
\end{table}

\subsection{SuperWASP Photometry}
\label{sec:swaspphot}

Time series photometry of WASP-12 was obtained in the 2004 and 2006 seasons 
with the SuperWASP-N camera located on La~Palma, Canary Islands \citep{swasp_instr}.  
In 2004, when the target was first observed, only 820 
photometric measurements were obtained between 2004 August and 2004 September.  
However, the same fields were observed again in 2006 after upgrades
to the telescope mount and instrument.  During the 2006 season, the
target was observed in the field-of-view of two separate cameras, and a total of 5573 photometric
brightness measurements were obtained between 2006 November and 2007 March.  The data obtained in
both seasons were processed with a custom built data reduction pipeline described in
\citet{swasp_instr}, and the resulting light curves were analysed using a modified
box least-squares algorithm \citep{hunter,kovacs} to search for the planetary transit signature.

The combined SuperWASP data showed a significant periodic dip in brightness with a 
period,  $P=1.091$~days, duration, $\tau\sim 2.7$~hours, and
depth, $\delta\sim 14$~mmag.  The improvement in $\chi^2$ of the box-shaped transit 
model over the flat light curve was $\delta\chi^2 = 719$ and 
the signal-to-rednoise \citep{Pont06} was SN$_{{\rm red}}=15.6$.  
A total of 23 partial or full transits were captured by SuperWASP.

\begin{figure}
  \centering
  \includegraphics[angle=0,width=\columnwidth]{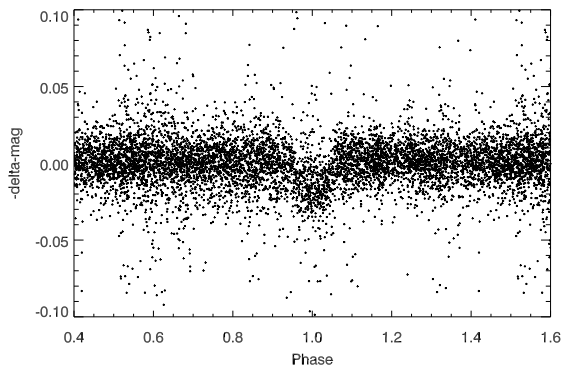}
  \caption{
   The SuperWASP discovery light curve of WASP-12 phase-folded with a period,  $P=1.09142$~days
   and epoch, $T_0 = 2454476.2321$.
  }
  \label{fig:swasp_lc}
\end{figure}

There were no obvious objects blended with WASP-12 in the SuperWASP aperture, and the detected transit 
event was significant, therefore,
WASP-12 was classed as a high priority target needing further study.  In Figure~\ref{fig:swasp_lc},
we show the phase-folded light curve of the SuperWASP data, adopting the ephemeris resulting from
the box-least squares analysis on the combined light curve.

\subsection{Follow-up Multi-band Photometry}

After identification as a high priority transit candidate, follow-up photometry
of WASP-12 was obtained using two additional telescopes with 
high spatial resolution ( $< 1^{\prime\prime}$/pixel) compared to 
SuperWASP ($13.7^{\prime\prime}$/pixel).  
We obtained observations of WASP-12 and the surrounding region during the 
predicted time of transit to confirm that there are no eclipsing binaries
within an arcminute of WASP-12 that may have caused of the transit
signal in the SuperWASP data.  
WASP-12 also appears to be a single star at the resolution provided by these data. 
The closest companion is $9^{\prime\prime}$ from the target and has a
magnitude of $V\sim18$.
Encouraged by this, we then obtained complete $B$, $I$, and $z$-band light
curves of the transit.  A detailed description of the follow-up photometry
is described below.

\subsubsection{Tenagra Telescope photometry}

WASP-12 was observed using the fully robotic, Tenagra II, 0.81m
f/7 Ritchey-Chretien telescope sited in Arizona, USA.  
The science camera contains a 1k$\times$1k SITe CCD with a 
pixel scale of $0.87^{\prime\prime}$/pixel and a field of view of 
$14.8^{\prime} \times 14.8^{\prime}$.  These photometric data were obtained
as part of an observing program sponsored by the Las Cumbres Observatory
Global Telescope Network\footnote{www.lcogt.net}.  
The $B$-band transit of WASP-12 was created from 227 observations on two consecutive 
nights in March.  The $I$-band light curve consists of 639 flux measurements taken over
five non-photometric nights in 2008 February and March.  

Calibration frames obtained automatically every twilight
were used in processing the images (bias subtraction and flat fielding)
with the SuperWASP data reduction pipeline \citep{swasp_instr}. 
Object detection and aperture photometry, with a 7.5~pixel radius aperture, were
performed on all the stars in the frame using DAOPHOT \citep{daophot}
run under IRAF\footnote{IRAF is written and supported by the IRAF
programming group at the National Optical Astronomy Observatories (NOAO)
in Tucson, Arizona.  NOAO is operated by the Association of Universities 
for Research in Astronomy (AURA), Inc.\ under cooperative
agreement with the National Science Foundation}.  The differential
photometry was derived from seven, non-variable comparison stars
within the field which had V$ < 14$~magnitude and J-H colors 
similar to the target star.
For each of the filters, the fluxes from the comparison stars were summed, 
and then converted into a magnitude which was then subtracted from the 
instrumental magnitude of the target star.   The resulting differential 
photometry of the $B$-band light curve has an rms of $\sim 7$~mmag.

The $I$-band transit was created from data obtained in
non-photometric conditions over five different visits to the
star (some at high airmass).  These data were essential in confirming the transit
and refining the ephemeris, but show significant red noise
systematics. Therefore, they were unsuitable for defining
accurate transit parameters and were excluded when modelling
the light curve.

\subsubsection{Liverpool Telescope photometry}

A full transit was observed using RATCam on the robotic 2m Liverpool Telescope (LT) \citep{liverpool}
on La Palma as part of the Canarian Observatories {\it International Time Programme}.
A total of 614 images were taken in the Sloan z' band on the night of 18 February 2008.
We employed 2$\times$2 binning (0.27$^{\prime\prime}$/pixel)
and used a 10s exposure time for all observations.  The night was photometric,
and the pointing was stable.  The target star did not drift more than a few
pixels throughout the transit which allowed for very accurate differential
photometry.

\begin{figure}
  \centering
  \includegraphics[angle=0,width=\columnwidth]{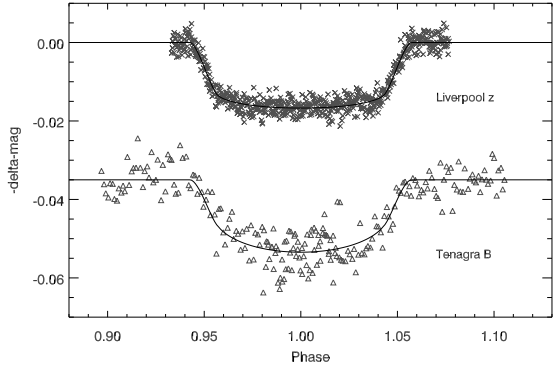}
  \caption{
   Differential $z$-band (top) and Tenagra $B$-band (bottom)
   photometry of WASP-12 during the transit.  An offset has been added
   to the $B$-band data for clarity.  The data are phase-folded
   with the ephemeris given in Table~\ref{tab:params}.  Overplotted are
   the best fit model transit light curves using the formalism
   of \citet{mandelagol} applying the limb darkening coeffs from \citet{claret00,claret04}.
  }
  \label{fig:followuplc}
\end{figure}

The images were debiased, flat-fielded, and corrected for fringing using
the standard RATCAM processing pipeline\footnote{see http://telescope.livjm.ac.uk/Info/TelInst/Inst/RATCam/}.
IRAF DAOPHOT was then used to obtain aperture photometry of the target
and four, non-variable nearby comparison stars using a 15~pixel radius aperture.  
The comparison stars were chosen primarily for brightness from within the 
limited FOV ($4.6^{\prime}\times4.6^{\prime}$) of
the camera and are not necessarily matched in color to the target.
The differential photometry was performed
in the same way as described above for the Tenagra data, and the
resulting out-of-eclipse light curve has 2.5~mmag precision.
The $z$-band and $B$-band transit
light curves obtained via follow-up photometry are plotted in Figure~\ref{fig:followuplc}.

\subsection{SOPHIE Spectroscopy}

High resolution spectroscopy of WASP-12 was obtained between 2008 February 12--22
with the SOPHIE spectrograph \citep{sophie} on the 1.93m telescope
at the Observatoire de Haute Provence.  We used the same 
observing program and instrument set-up as for other SuperWASP planets 
discovered in the northern hemisphere \citep{wasp3,wasp10,wasp11}.

The weather was clear and reasonably stable throughout the run, although
there were some fluctuations in seeing and transparency.  Therefore, we adopted
exposure times of 900s and 1080s depending on the transparency.
This allowed us to achieve a signal-to-noise (S/N) per resolution element (at 5500\AA)
of 30-45 in all the spectra.  The S/N values for all the spectra
are listed in Table~\ref{tab:logspec}.  The spectra were processed in real time
with the SOPHIE instrument control and data reduction software, 
and the RV measurements were obtained using a weighted cross-correlation
method \citep{Baranne,Pepe}.  To do this, we used a 
numerical mask constructed from the solar spectrum atlas corresponding to a G2V star.

The Moon was bright over most of the run thus contaminating the spectra
with scattered light.  We removed its velocity 
signature (according to the procedure described in \citet{wasp3}) from 
all the spectra.  We then fit the resulting cross-correlation function from each spectrum
with a Gaussian to derive a value for the central RV,
the full width half maximum (10.15-10.30 km~s$^{-1}$), and the peak amplitude or Contrast
(28.60--29.44\%).  The uncertainties for all RV measurements were
derived using an empirical relation applicable to SOPHIE spectra 
taken in the high efficiency mode  \citep{wasp11,bouchyinprep}.

\begin{figure}
 \centering
 \includegraphics[angle=0,width=\columnwidth]{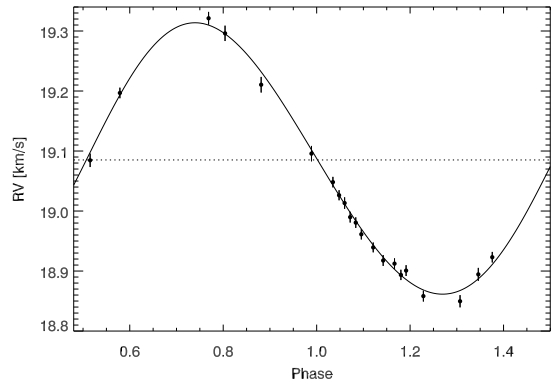}
 \caption{The SOPHIE radial velocity curve of WASP-12 phase-folded with
 the ephemeris given in Table~\ref{tab:params}.  The solid line is
 the best model curve resulting from the orbital parameters
 of the system derived from the MCMC analysis in \S\ref{sec:mcmc}.
 The systemic RV value is shown by the dotted line.
 }
 \label{fig:rvcurve}
\end{figure}

In this way, we obtained 21 RV measurements of WASP-12 over ten nights
which have typical uncertainties of $\sim$~10~m~s$^{-1}$.  The RV measurement
obtained from each spectrum is listed in Table~\ref{tab:logspec} along with the
derived uncertainty.  These data have a
standard deviation of 130~m~s$^{-1}$, significantly greater than the individual
uncertainties, and they vary sinusoidally when folded
on the period derived from the transit photometry (Figure~\ref{fig:rvcurve}).

\begin{table}
\caption[]{The radial velocity measurements of WASP-12 obtained with SOPHIE spectrograph.  }
\label{tab:logspec}
\begin{center}
\begin{tabular}{ccccc}
BJD   & $\rm{V}{r}$  & $\rm{\sigma_{RV}}$  & S/N & Bisector \\
     &  km s$^{-1}$ & km s$^{-1}$          &      & km s$^{-1}$\\
\hline\\
2454509.38633   &    18.9231  &   0.0088 & 44.3 &  0.015    \\
2454509.53836   &    19.0845  &   0.0112 & 34.9 &  0.047  \\
2454510.40255   &    18.8497  &   0.0102 & 38.0 &  0.046   \\
2454511.29105   &    18.9394  &   0.0082 & 47.4 &  0.056   \\
2454511.36791   &    18.9008  &   0.0090 & 43.0 &  0.047   \\
2454511.40825   &    18.8582  &   0.0090 & 43.3 &  0.047   \\
2454511.53661   &    18.8945  &   0.0108 & 36.1 &  0.005   \\
2454512.28835   &    19.0648  &   0.0088 & 44.0 &  0.057   \\
2454512.30278   &    19.0429  &   0.0084 & 45.6 &  0.031   \\
2454512.31570   &    19.0298  &   0.0096 & 40.1 &  0.010   \\
2454512.32867   &    19.0064  &   0.0090 & 42.7 &  0.033   \\
2454512.34174   &    18.9973  &   0.0088 & 44.2 &  0.030   \\
2454512.35470   &    18.9778  &   0.0086 & 45.2 &  0.036   \\
2454512.40579   &    18.9176  &   0.0090 & 42.6 &  0.038   \\
2454512.43225   &    18.9124  &   0.0088 & 44.4 &  0.047   \\
2454512.44721   &    18.8936  &   0.0086 & 45.6 &  0.018   \\
2454513.32972   &    19.0957  &   0.0128 & 29.7 &  0.079   \\
2454514.30357   &    19.2105  &   0.0130 & 29.6 &  -0.001  \\
2454515.27220   &    19.3213  &   0.0106 & 36.1 &  -0.018  \\
2454516.40210   &    19.2961  &   0.0128 & 29.8 &  0.067   \\
2454519.43071   &    19.1968  &   0.0088 & 44.3 &  0.038   \\
\hline\\
\end{tabular}
\end{center}
\label{default}
\end{table} 

\begin{figure}
 \centering
 \includegraphics[angle=90,width=\columnwidth]{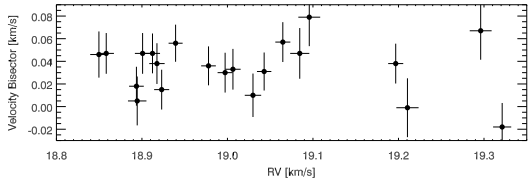}
 \caption{Line-bisector velocity versus radial velocity measured
  from all the observed SOPHIE spectra.  We adopt uncertainties
  of twice the radial velocity uncertainty for all bisector measurements.
  There is no correlation between these two parameters indicating 
  the radial velocity variations
  are not caused by stellar activity or line-of-sight binarity.
 }
 \label{fig:bisec}
\end{figure}

Finally, any asymmetries in the line profiles were explored by measuring the
velocity span of the line-bisector \citep{gray} according to the techique outlined in \citet{bisec2}.
These measurements, also listed in Table~\ref{tab:logspec},
show no correlation with radial velocity (Figure~\ref{fig:bisec}), therefore it is unlikely that
the observed RV variations are caused by star spots on the stellar surface or by 
blending with an eclipsing binary star in the system or close to the line-of-sight.

Thus, we conclude that the observed RV variations 
are caused by the gravitational influence of a planetary-mass object
orbiting WASP-12.  Figure~\ref{fig:rvcurve}
shows a plot of the RV measurements phase-folded on the ephemeris
given in Table~\ref{tab:params} and overplotted with the best fitting
model radial velocity curve which is derived as described in
the analysis section (\S\ref{sec:mcmc}).

\subsection{Additional Spectroscopy}
\label{sec:addspec}

The SOPHIE spectra were obtained in the high efficiency (HE) mode which
is known to suffer from problems with corrections for the blaze shape.
According to the data products website, a residual blaze effect remains in
the reduced spectra at the 5\% level\footnote{see http://www.obs-hp.fr/www/guide/sophie/data\_products.html}.
These data provide exceptional radial velocity precision, but
they are not always suitable for determining stellar parameters
For example, the effective temperature is strongly constrained
by the wings of the H$\alpha$ line, and for hot stars
with broad H$\alpha$ wings, the residual instrumental features
in the SOPHIE HE spectra lead to large uncertainties
on this parameter.  Therefore, in addition to the 21 SOPHIE 
spectra which were used primarily to measure the radial velocity 
signature of the orbiting planet, several additional spectra of WASP-12 were 
obtained and used in deriving independent 
measurements of the stellar parameters of the host star.
The additional spectroscopic observations are described below, and the derivation
of the stellar parameters from all available spectra are described
in \S\ref{sec:specsynth}.

\subsubsection{Isaac Newton Telescope spectra}

Two individual spectra of WASP-12 were obtained on 2008 April 22 with the 2.5m~Isaac
Newton Telescope and Intermediate Dispersion Spectrograph (IDS).  The longslit data
were taken with the H1800V grating using a 1.2$^{\prime\prime}$ slit which
resulted in a resolution of ($R\sim 8000$).  A signal-to-noise
of $> 50$ was achieved in both individual spectra by taking 900s exposures.
The spectrum was centered at 6500\AA\ and covered the region from 6200--7000\AA,
thereby providing measurements of H$\alpha$, the Li~I doublet at 6708\AA\ and many narrow metal lines.
Biases and lamp flats were obtained at the beginning of the night and Neon-Copper-Argon
arcs were taken just before and just after the WASP-12 observations.
The spectra were reduced in a standard way using the IRAF~{\it longslit} package.
We then averaged the two individual spectra and
continuum normalized the composite observation before fitting for the
stellar parameters.

\subsubsection{Telescopio Nazionale Galileo spectra}

We observed the target again on 29 April 2008 using the high efficiency echelle spectrograph, 
SARG, mounted on the 3.58~m~Telescopio Nazionale Galileo (TNG) 
telescope. These data were taken as part of the Canarian Observatories {\it International Time Programme}.
Three consecutive 1800 second exposures were taken using the yellow filter and grism.
The spectra were binned 2$\times$1 in the spatial direction at the time of observation to reduce the readout time.
A slit width of 0.8$^{\prime\prime}$ was adopted which resulted in a spectral resolution of R$\sim$57000.
Calibration images, including bias frames, lamp flat-field frames, and Thorium-Argon arcs, were
taken at the beginning of the night and used in processing the target spectra with the {\sc reduce}
echelle data reduction package \citep{pv2002}.  Special care was taken to provide an accurate flat-fielding
of the data.  The three individual reduced spectra were averaged on an order-by-order basis
to produce a final merged spectrum which was then used in the determination
of the stellar parameters described in \S\ref{sec:specsynth}.

\section{Analysis}
\label{sec:analysis}

\subsection{Spectroscopic analysis}
\label{sec:specsynth}

Three spectra of WASP-12 were derived from observations with the SARG,
SOPHIE, and IDS spectrographs.  
Each independent spectrum was compared with synthetic
spectra to determine the effective temperature $T_{\rm{eff}}$,
gravity ${\rm log}\,g$, metallicity, [M/H], and projected
stellar rotation $v\,{\rm sin}\,i$ of WASP-12. 
Our spectral synthesis technique closely follows the 
procedure of \citet{v2} (hereafter VF05), and
a detailed description can be found in \citet{eric}.

Two additional parameters, microturbulence and macroturbulence, are incorporated
into the spectral synthesis to characterise turbulent 
mixing and convection in the upper layers of stellar atmospheres \citep{gray}.
Their chosen values affect the derived stellar properties such that
microturbulence anti-correlates strongly with metallicity, and
macroturbulence affects the line broadening, and therefore the $v\,{\rm sin}\,i$ measurement.
In our spectral synthesis, we closely follow VF05, so that
our results can be compared directly with this extensive
spectroscopic analysis of planet-hosting stars.  For
the microturbulence, we adopt their value of $v_{mic} =0.85$~km~s$^{-1}$,
but we note that other empirical studies of main sequence F-stars 
suggest higher values for $v_{mic}$ \citep[see][]{montalban}.
For the macroturbulence, we use the empirical linear relation with
temperature provided in VF05 to derive a value
of $v_{mac}=4.8$~km~s$^{-1}$.  However, larger values are not excluded 
for hot stars like WASP-12.  This is due to the 
difficulty in accurately measuring macroturbulence for early type stars
where rotational broadening typically dominates the line widths.
Furthermore, only 79 stars in the VF05 sample have T$_{eff}\ge 6200$~K, so
the empirical relation is not very well defined in this regime.

\begin{figure*}
  \includegraphics[angle=90,width=\textwidth]{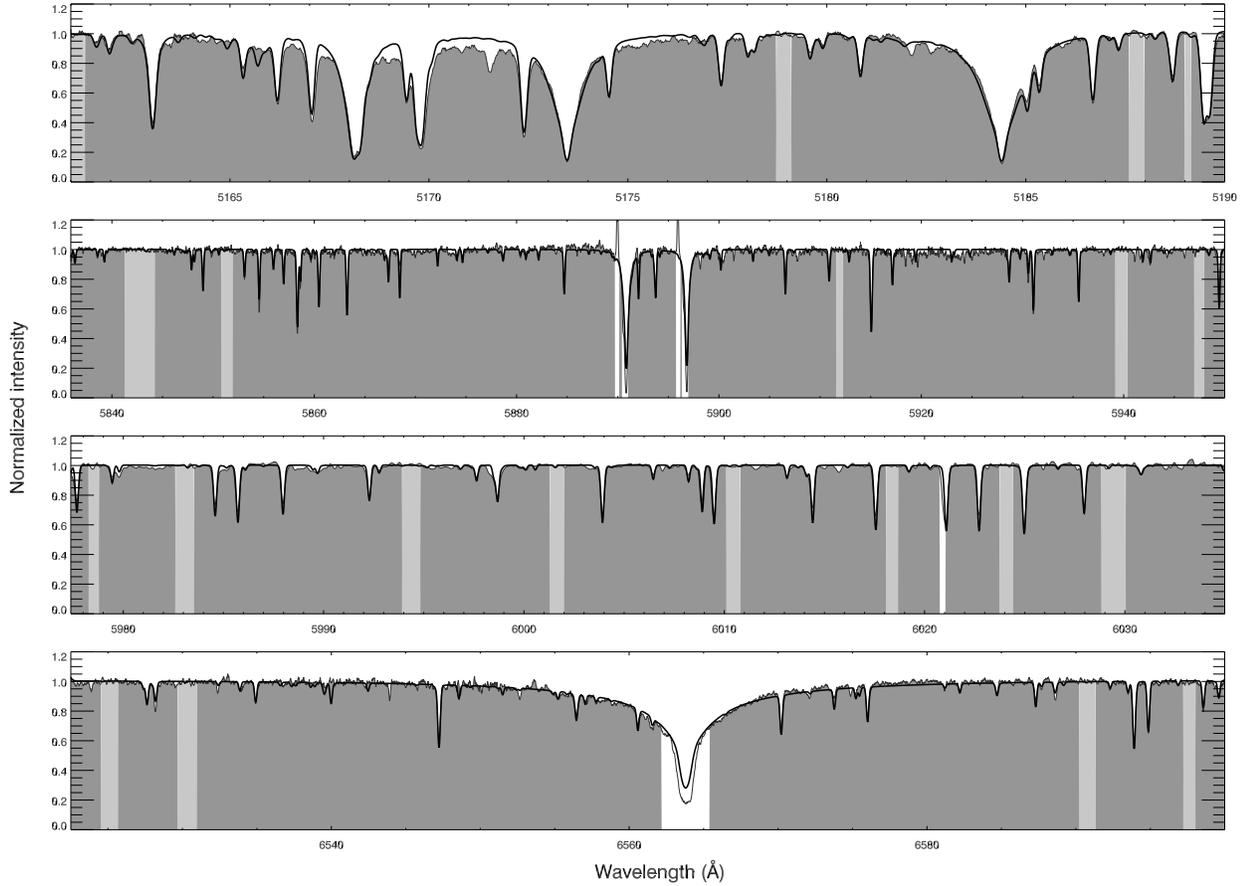}
  \caption{Observed SARG spectrum (grey) overplotted by the
   best fitting theoretical model spectrum (solid black line).  The
   top panel shows the region around the Mg~b triplet (5160-5190\AA), 
   the second panel is the region around the Na~I~D doublet (5850-5950\AA), 
   the third panel shows the region from 6000-6210\AA\ with
   a large number of metal lines, and the bottom panel is the region 
   around the H$\alpha$ line (6520-6600\AA).  These regions are modelled
   simultaneously with spectral synthesis to derive the parameters
   of the host star.  The light grey regions of the spectrum
   are used to determine the continuum.  Note, narrow telluric emission features
   are present in second panel at the rest wavelength of the Na I doublet feature.}
  \label{fig:specsynth}
\end{figure*}

We used the SARG data to derive our best 
measurement of the parameters of the host star. This spectrum is 
of high resolution and good quality with no known residual instrumental features.
Four spectral regions (shown in Figure~\ref{fig:specsynth}) 
were selected for the fit because they are particularly sensitive 
to one or more of the parameters we aim to derive.  
A simultaneous fit to these four regions of the spectrum yielded a 
$T_{\rm{eff}}=6290$ K, ${\rm log}\,g = 4.38$, and  [M/H]=$0.30$.
The line broadening was equivalent to that of the spectral resolution,
therefore we were only able to derive an upper limit on the 
rotational broadening of $v\,{\rm sin}\,i < 2.2$~km~s$^{-1}$.  This is derived by
subtracting (in quadrature) the estimated value of macrotubulence
(4.8 km~s$^{-1}$) from the width of the smallest resolvable resolution element
(5.3 km~s$^{-1}$ at R=57000).  
A comparison of the observed and best fitting model spectrum
is shown in Figure~\ref{fig:specsynth}, and
the final stellar parameters are listed in Table~\ref{tab:specparms}.   We 
derive uncertainties on these properties based on the range of values
measured from the additional analysis of the IDS and SOPHIE data.

In the SOPHIE spectrum, we simultaneously fit the same four
spectral regions given above and find 
$T_{\rm{eff}}=6175$~K, ${\rm log}\,g = 4.36$, and [M/H]=$0.15$.
We also measure a $v\,{\rm sin}\,i = 4.5$~km~s$^{-1}$. Again, 
we believe this value is only an upper limit.  
The IDS spectrum spanned the region of H$\alpha$ and surrounding
metal lines.  In this region,
there are no strongly gravity-sensitive features, therefore, we
fixed the ${\rm log}\,g = 4.36$, which was determined from the SOPHIE
observations, and solved for the stellar temperature, $T_{\rm{eff}}=6495$~K,
and metallicity, [M/H]$=0.16$.  The resolution was too low to
measure $v\,{\rm sin}\,i$.  

In summary, we made three independent measurements of the properties of WASP-12
by comparing spectroscopic observations of the star to 
model spectra.  In all three analyses, we find that WASP-12 is a hot, slowly rotating, 
metal rich, dwarf star.  We adopt the results from the analysis of the SARG spectrum as
our final values for the parameters of WASP-12 and the uncertainties
on the stellar parameters from the range of values that were determined in 
the three different analyses.  

\subsection{Deriving Planet and Host Star Parameters}
\label{sec:mcmc}

The multi-band light curves and radial velocity curve of WASP-12 were
analysed simultaneously in a Markov-chain Monte Carlo (MCMC) based
routine designed specifically to solve the multivariate problem of
transiting star-planet systems.  The routine is described in detail
in \citet{mcmc} and \citet{wasp3}.  
The results of the box-least squares analysis to the SuperWASP
photometric data (described in \S~\ref{sec:swaspphot}) provide an initial estimate of the
light curve parameters.  We also initially assume an eccentricity
of 0.02, a systemic RV equal to the mean of the velocity data, and a velocity
amplitude derived by fitting a sinusoidal velocity variation to the observed
RVs by minimizing $\chi^2$.  To derive a first 
guess for the stellar mass, we interpolate the super-solar metallicity
(Z=0.03, [M/H]=0.2), zero-age main sequence temperature-mass relation from
\citet{girardi00} at the stellar temperature derived in the previous
section, $T_{\rm{eff}}=6300$~K.  We adopt a stellar mass of 1.28~M$_{\odot}$
as the initial value for this parameter.

Via the MCMC approach, the routine repeatedly adopts trial parameters until
it converges on the set of values which produce the best model velocity 
curve and model light curves.  The model light curves
are generated from the analytic transit formulas found in \citet{mandelagol}
(adopting the small-planet approximation) and using the limb darkening
coefficients for the appropriate photometric filters from \citet{claret00,claret04}.
The sum of the $\chi^2$ for all input data curves with respect to the 
models is the statistic used to determine goodness-of-fit.  The routine
also produces $1\sigma$ uncertainties on all the parameters.  

\subsubsection{Evolutionary Status of the Host Star}

We ran the MCMC code initially using the SuperWASP light curve, the 
$B$ and $z$-band follow up photometry, and the SOPHIE radial velocity curve as input data sets.
We ran the code without imposing
the main sequence prior on the overall $\chi^2$ statistic.
This is not unreasonable for a late-F star which, according to the theoretical models,
has a main sequence lifetime of $\la 1$~Gyr.
We then determined the evolutionary status of the host star using the 
results of this run to assess whether this was a reasonable assumption.

First, we examined the lithium abundance in WASP-12 as a possible
age indicator.  In the SARG spectrum, there is no absorption detected in the Li~I line
located at 6708\AA.  The IDS spectrum shows a broad, very shallow 
absorption feature at this position,
however due to the lack of detection in the SARG spectrum, we suspect
the feature is due to noise or blending with other absorption lines (e.g.\ Fe~I at 6707.44\AA).
This lack of Li is consistent with low levels found in old open clusters, like M~67 ($\sim 4$~Gyr),
for a 6300~K star \citep{sesitorandich},
however a precise age determination cannot be derived for the star from this observation.

\begin{figure}
  \centering
  \includegraphics[angle=0,width=\columnwidth]{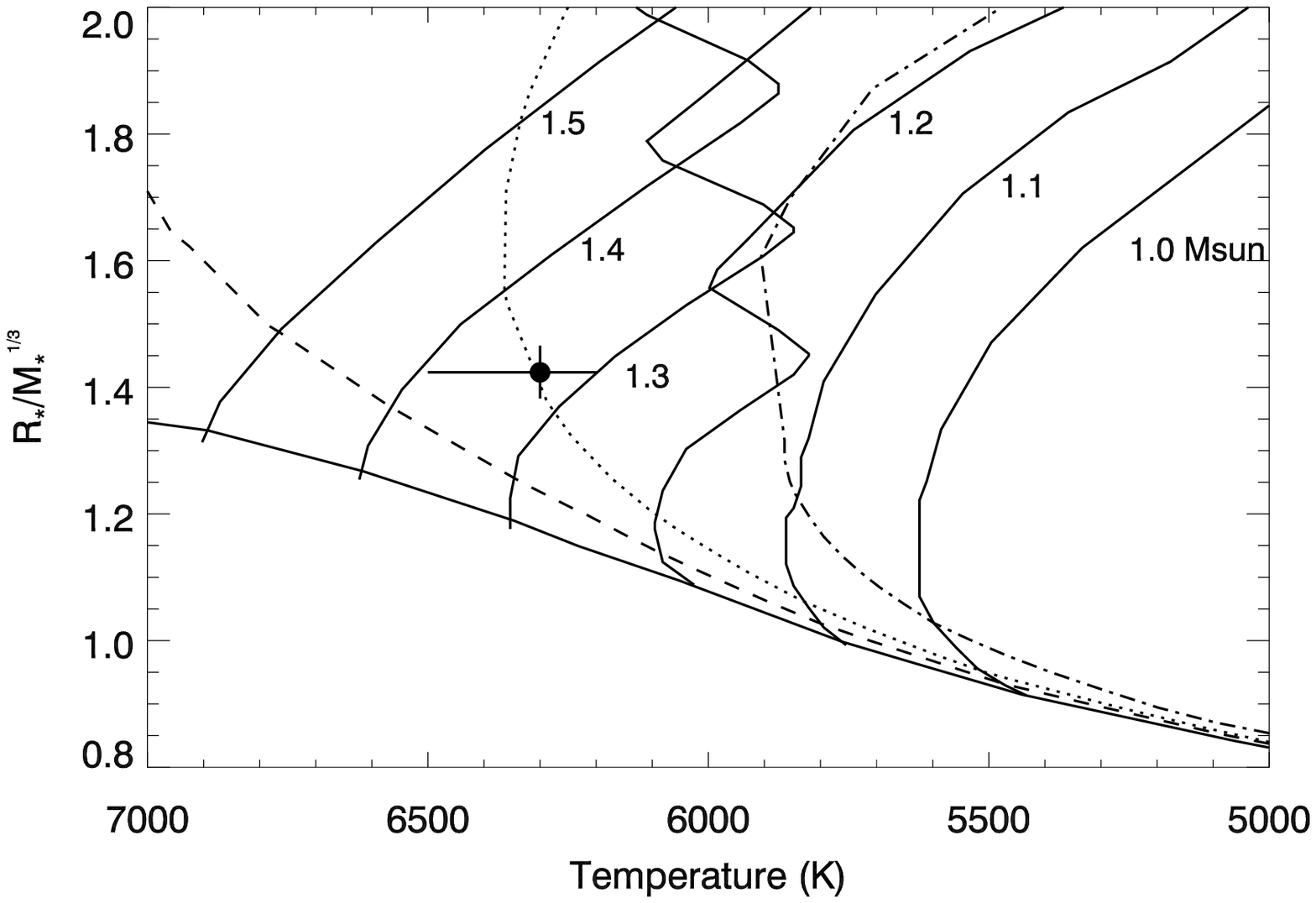}
  \caption{$R_{*}/M_{*}^{1/3}$ in solar units versus effective temperature. WASP-12 is the large solid circle.
   Evolutionary mass tracks (solid lines with adjacent numbers labelling the
   mass of that line) and isochrones (100~Myr (solid), 1~Gyr (dashed), 2.0~Gyr (dotted), 5.0~Gyr (dot-dashed)
   from \citet{girardi00} for Z=0.03 are plotted for comparison to the WASP-12 parameters.  }
  \label{fig:hrplot}
\end{figure}

Next, we compare the structure and temperature of the star to the super-solar metallicity stellar 
evolution models of \citet{girardi00} to constrain the age.  We use the
[M/H]=0.2 (Z=0.03) tracks which are consistent with the measured metallicity of WASP-12,
given the uncertainty on this parameter.
Figure~\ref{fig:hrplot} shows a modified Hertzprung-Russel diagram comparing the host star to 
the theoretical mass tracks and isochrones.  Here, we plot the inverse cube root
of the stellar density, $R_{*}/M_{*}^{1/3}$, in solar units 
versus the stellar temperature.    
We choose to compare the data to the models in this parameter space since the 
quantity, $R_{*}/M_{*}^{1/3}$, unlike $R_{*}$ or luminosity, 
is purely observational and is measured directly from the light curve.  
In addition, it is completely independent of the temperature determined from the spectrum.  

The results of the initial MCMC run provided
a measurement of the mean stellar density.  We converted the density to $R_{*}/M_{*}^{1/3}$ in solar
units, and generated the same property from the mass and log~$g$ values in the models. We then interpolated
in the $R_{*}/M_{*}^{1/3}-T_{\rm eff}$ plane to determine the mass and age for WASP-12.
We interpolated linearly along two consecutive mass tracks to generate an equal number of age points
between the zero-age main sequence and 
the coolest point at the end of core hydrogen burning.  We then interpolated between the mass tracks along
equivalent evolutionary points to find the mass and age from the models that best match
the stellar properties derived from the MCMC code and the spectral snythesis.
According to these particular tracks, the large value for $R_{*}/M_{*}^{1/3}$ 
indicates the star has evolved off the zero age main sequence, but has yet to reach the shell hydrogen burning stage.  
It is in a position in the diagram which give it a mass, $M_{*}=1.33\pm0.05$~M$_{\odot}$ 
and an age of $2.0^{+0.5}_{-0.8}$~Gyr.  To check the accuracy and precision of this result, we 
compared the star to a second set of stellar
evolution models.  When interpolating the Z=0.03 tracks by \citet{yi}, we find a similar
result.  The position of the star in the $R_{*}/M_{*}^{1/3}-T_{\rm eff}$ gives
an age of 2.4~Gyr and a mass of $1.38$~M$_{\odot}$.  

We investigated using the rotation period of the star which allows for constraining
the age based on the expected spin-down timescale.
The slow rotational period argues for an old age, however with only an upper limit 
on the $v\,sin\,i$ we are unable derive an age estimate using the gyrochronology 
technique \citep{barnes}.

The three age-dating techniques discussed all suggest WASP-12 is several Gyr old,
however the stellar evolution models provide the only definitive estimate of the age.
Therefore we adopt a final age for WASP-12 of $\tau = 2\pm 1$~Gyr.
We have increased the uncertainty to include the error in metallicity and the
systematics in the stellar evolution models.

\subsubsection{Final Determination of Planet and Host Star Parameters}

After determining the evolutionary status of the host star and
estimating the stellar mass from the evolutionary models,
we ran the MCMC code a second time.  We did not impose the main sequence
mass-radius prior, since the star has evolved off the zero age main sequence.
Furthermore, we adopted an initial estimate for the stellar mass of 1.33~M$_{\odot}$
from the comparison to the theoretical tracks.  

In addition, during the radial velocity observing run, we attempted to detect the Rossiter-McLaughlin effect
by taking six consecutive measurements of the target during and just after the expected transit.  
We see no evidence for this effect in the data which confirms a small $v\,sin\,i$ for WASP-12, and
is consistent with the upper limit on the projected rotational velocity derived from the spectral synthesis.  
We do not currently use the in-transit radial velocity data to
derive an independent measurement of $v\,sin\,i$, however we do not
wish for this sequence of observations, comprising 1/4 of our RV measurements, 
to bias the fit to the system orbital parameters.  
A small, but systematic zero-point offset in these data could affect the final results.
Therefore, we treated the six data points taken in sequence on
the same night as a different dataset during the fitting which allows them 
to have a slight zero-point offset compared to all the other data points. 
The resulting zero point offset is only 0.014~m~s$^{-1}$.
 
The results of the final fit to the light curve and radial velocity curve 
using the SOPHIE radial velocity data, the SuperWASP photometry, the Liverpool Telescope 
$z$-band, and the Tenagra $B$-band data are given in Table~\ref{tab:params}
with their $1\sigma$ uncertainties.  The model light curves in each band are overplotted 
on the phase-folded data in Figure~\ref{fig:followuplc}, and the model
radial velocity curve is shown in Figure~\ref{fig:rvcurve}.  
In the analysis, we find that WASP-12b is a low density
planet ($\sim0.2\rho_J$) with a mass approximately 40\% larger than Jupiter in a 1~day orbit
around a hot, metal rich late-F star which is evolving off the 
zero age main sequence.

We closely examined the most fundamental fitted parameters (depth, width, and impact parameter)
which are used to describe the shape of the transit itself and find the
model to be a good match to the data. 
We do note that the impact parameter, $b$, which is determined by the shape
of the ingress and egress, is difficult to measure precisely without exceptionally
good photometry.  Furthermore, this parameter strongly influences the stellar
radius and thus the planet radius through the depth measurement.
Since the impact parameter has such a subtle effect on the light
curve, but a large effect on the resulting parameters, 
we calculate an absolute lower limit on the planet radius by running the MCMC
code with the impact parameter fixed to zero (b=0 model).  
The b=0 model has a greater $\chi^2$ compared to the 
overall best fitting model by 15, which is significant, 
but the effect on the light curve is very subtle.
However, when comparing the results of the b=0 model to the overall best fit,
we find a stellar radius of $R_{*\_{\rm b=0}}=1.46$~R$_{\odot}$ 
compared to $R_{*}=1.57$~R$_{\odot}$ and a planet radius of
$R_{\rm pl\_b=0}=1.63$~R$_{\rm J}$ compared to $R_{\rm pl}=1.79$~R$_{\rm J}$.



\begin{table*}
\begin{center}
\caption[]{WASP-12 system parameters and 1$\sigma$ error limits derived
from the MCMC analysis.}
\label{tab:params}
\begin{tabular}{lccl}
\hline\\
Parameter & Symbol & Value & Units \\
\hline\\
Transit epoch (BJD) & $ T_0  $ & $ 2454508.9761^{+ 0.0002 }_{- 0.0002 } $ & days \\
Orbital period & $ P  $ & $ 1.091423^{+ 0.000003 }_{- 0.000003 } $ & days \\
Planet/star area ratio  & $ (R_p/R_s)^2 $ & $ 0.0138^{+ 0.0002 }_{- 0.0002 } $ &  \\
Transit duration & $ t_T $ & $ 0.122^{+ 0.001 }_{- 0.001 } $ & days \\
Impact parameter & $ b $ & $ 0.36^{+ 0.05 }_{- 0.06 } $ & $R_*$ \\
  &    &      &  \\
Stellar reflex velocity & $ K_1 $ & $ 0.226^{+ 0.004 }_{- 0.004 } $ & km s$^{-1}$ \\
Centre-of-mass velocity  & $ \gamma $ & $ 19.0845^{+ 0.002 }_{- 0.002 } $ & km s$^{-1}$ \\
Orbital semimajor axis & $ a $ & $ 0.0229^{+ 0.0008 }_{- 0.0008 } $ & AU \\
Orbital inclination & $ I $ & $ 83.1^{+ 1.4 }_{- 1.1 } $ & degrees \\
Orbital eccentricity & $ e $ & $ 0.049^{+ 0.015 }_{- 0.015 } $ &  \\
Longitude of periastron & $ \omega $ & $ -74^{+ 13 }_{- 10 } $ & deg  \\
  &    &      &  \\
Stellar mass & $ M_* $ & $ 1.35^{+ 0.14 }_{- 0.14 } $ & $M_\odot$ \\
Stellar radius & $ R_* $ & $ 1.57^{+ 0.07 }_{- 0.07 } $ & $R_\odot$ \\
Stellar surface gravity & $ \log g_* $ & $ 4.17^{+ 0.03 }_{- 0.03 } $ & [cgs] \\
Stellar density & $ \rho_* $ & $ 0.35^{+ 0.03 }_{- 0.03 } $ & $\rho_\odot$ \\
  &    &      &  \\
Planet radius & $ R_p $ & $ 1.79^{+ 0.09 }_{- 0.09 } $ & $R_J$ \\
Planet mass & $ M_p $ & $ 1.41^{+ 0.10 }_{- 0.10 } $ & $M_J$ \\
Planetary surface gravity & $ \log g_p $ & $ 2.99^{+ 0.03 }_{- 0.03 } $ & [cgs] \\
Planet density & $ \rho_p $ & $ 0.24^{+ 0.03 }_{- 0.02 } $ & $\rho_J$ \\
Planet temperature ($A=0$,F=1)  & $ T_{\mbox{eq}} $ & $ 2516 ^{+ 36 }_{- 36 } $ & K \\
\hline\\
\end{tabular}
\end{center}
\end{table*}

\section{Discussion}
\label{sec:discuss}

\begin{figure}
  \centering
  \includegraphics[angle=0,width=\columnwidth]{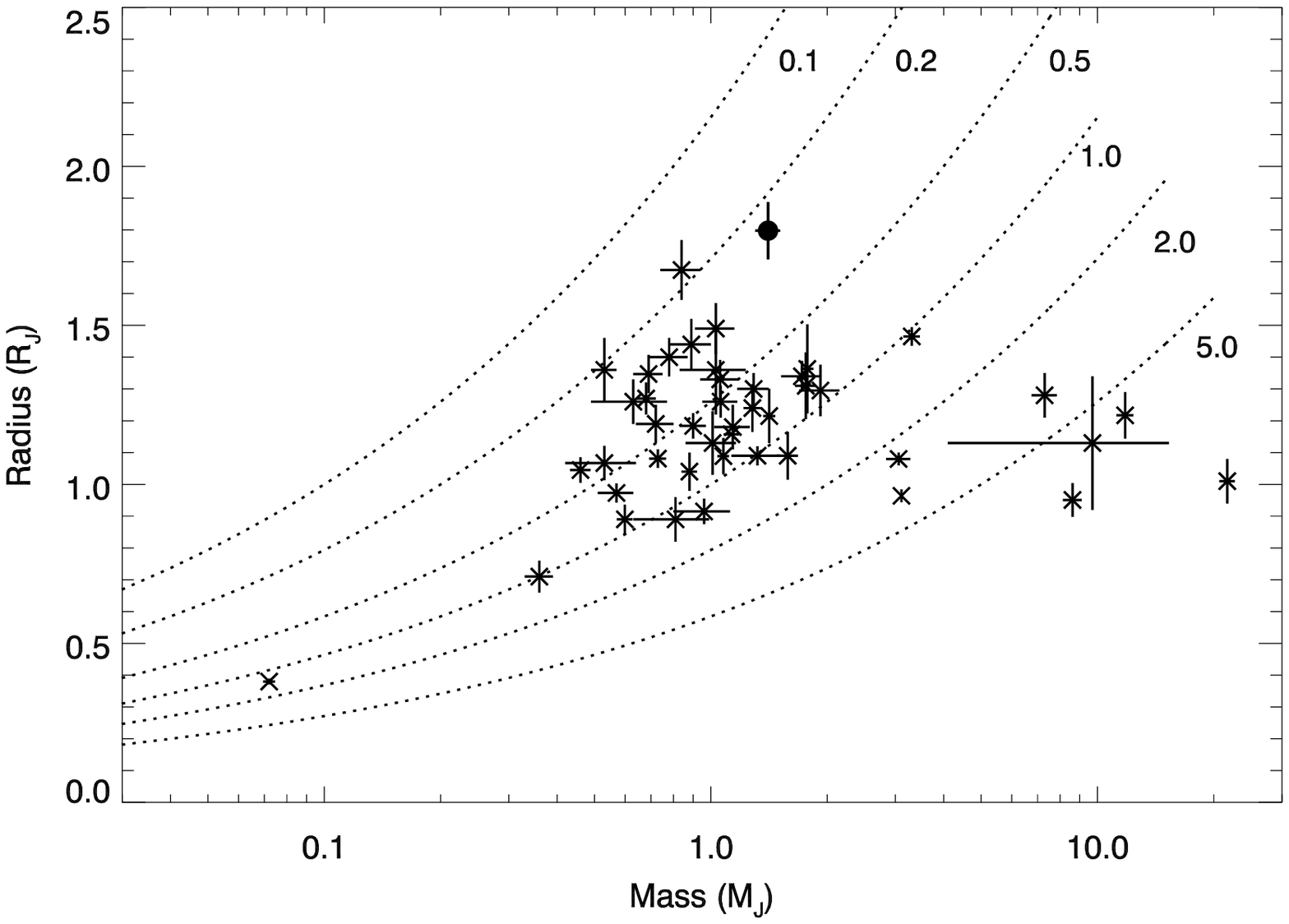}
  \caption{Mass versus Radius of all transiting planets 
   which have published masses and radii.  The data
   were obtained from the exoplanet encyclopedia (http://exoplanet.eu/).
   WASP-12b is the solid black circle.  Lines of constant planet density
   are overplotted (dotted) in units of Jupiter density.
  }
  \label{fig:mrplot}
\end{figure}

WASP-12b is a unique transiting planet with the most apparent feature
being the very large observed radius ($1.79$~R$_{J}$).  The planet has
a mean density only 24\% that of Jupiter, making it a
member of the growing class of transiting gas giants which all have particularly large radii.  
Figure~\ref{fig:mrplot} shows the position of WASP-12b among the other published
transiting planets in the mass-radius plane.
The structure of these planets, including HD~209458b  and WASP-1b,
cannot be explained through simple, isolated planet formation models, and a
great deal of recent theoretical work has gone into determining the mechanism or
mechanisms causing their large sizes.  It is clear that the external
environment (stellar irradiation, tidal forces), the internal properties 
(heavy element abundance, clouds/hazes, day-night heat transfer, core mass), and 
the evolutionary state (age) can all affect a planet's radius \citep{guillotshowman,boden03,fortney07,
burrows07,fortney08,burrows08a,burrows08b,jackson08}.

WASP-12b is the most heavily irradiated planet yet detected.
With a host star luminosity, L=3.48~L$_{\odot}$, the planet experiences
an incident flux at its substellar point of F$_P$=9.03$\times10^9$~ergs~cm$^{-2}$~s$^{-1}$.
which is twice the stellar flux experienced by OGLE-TR-56b or OGLE-TR-132b,
the next most highly irradiated planets \citep{torres,burrows07}.
This intense stellar radiation results in an equilibrium temperature,
T$_{eq}$=2516~((1-A)/F)$^{0.25}$~K, where A is the fraction of absorbed flux
and F is the fraction of the planet's surface that emits at T$_{eq}$.  Although
Jupiter has an absorbing fraction, A$=0.28$ \citep{jupalbedo}, existing evidence suggests
hot Jupiters have much lower albedos. For example, high precision optical photometry of HD~209458b 
gives an geometric albedo of only 4\% \citep{Rowe07}.

It is clear from the detailed modelling of hot Jupiters \citep{boden03,fortney07,burrows07} that
increased incident stellar radiation will lead to an increase in the planet radius
by inhibiting its contraction.  This is a function of stellar
mass and age, and less massive planets are more affected by the star's radiation in this
regard.  The extremely irradiated case 
of the relatively massive WASP-12b planet ($M_p$=1.41 M$_J$) at an age of $\sim 2$~Gyr is
outside the range of situations presented in these papers,
however, \citet{fortney07} give a radius of 1.248~R$_J$ for an object similar to WASP-12b
(1~Gyr old, 1.46~M$_{J}$), but at a distance of 0.02~AU from a solar luminosity star.  
In the context of the models, the incident radiation from WASP-12 would be equivalent to 
putting this simulated planet at a distance of $\sim0.01$~AU from a solar luminosity star.
Despite WASP-12b being more intensely irradiated than the simulated planet of the same age and
mass, it is difficult to see how stellar irradiation alone could result in the observed 
radius of $R_{\rm pl}=1.79$~R$_J$ (or even the absolute minimum radius of $R_{\rm pl_{\rm b=0}}=1.63$~R$_J$) 
in a planet as massive as WASP-12b.  

\citet{burrows07} show how enhancing the atmospheric opacities of extrasolar planets
also results in increased radii by maintaining the heat and entropy stored in the cores 
over longer timescales.  In their model, metallicity is used as a proxy for 
atmospheric heavy element abundance and thus opacity.  Employing an abundance of 10$\times$
solar allows the authors to fit the radii of the largest planets without adding any
additional internal heat sources.
It is very likely that WASP-12b has a super-solar atmospheric heavy
element abundance, given the super-solar metallicity of the host star.
Furthermore, according to the models of \citet{fortney08}, WASP-12b should be in the 
class of highly irradiated, hot planets (pM class) in which high-altitude molecular hazes
of TiO and VO (condensation temperature, T=1670~K) absorb strongly
at optical wavelengths resulting in larger radii.  
For WASP-12, measuring the primary and secondary eclipse depths as a function of 
wavelength and obtaining precise
out-of-eclipse photometry will allow for investigating the presence of a high
altitude absorbing population in the planet's atmosphere. 

Although \citet{burrows07} do not calculate a model specifically for WASP-12b, 
they are able to reproduce a $1.30$~R$_J$, $1.29$~M$_J$, 2.5~Gyr old planet (OGLE-TR-56b).
It will be interesting to see if a complete modelling of the
extreme environmental conditions of WASP-12b including stellar irradiation, increased heavy element
abundance, and high altitude hazes can produce a sufficiently 
large planet radius to match the observed value.  

If WASP-12b's large  radius cannot be explained by increasing the 
atmospheric opacity or other atmospheric phenomena, an additional source of 
internal energy will be required to explain the observations.  Dissipation 
of tidal energy is a possible contributor.  The timescale for
circularization of the WASP-12b orbit is most likely very short, 
but the best model fit to the observed radial
velocity and light curves produces a non-zero eccentricity of $e=0.049^{+ 0.015 }_{- 0.015 }$.  
Using the formalism for the circularization timescale taken 
originally from \citet{goldsoter66} and provided in \citet{boden03},
we find $$\tau_{\rm{circ}} = 3.2\frac{Q_P}{10^6}~{\rm Myr}$$ which
is much shorter than the 2~Gyr age of WASP-12, when $Q_P$
the tidal dissipation constant for the planet, is given the nominal
value of $10^6$.  

The non-zero eccentricity detection is barely
a 3$\sigma$~result, however, if after further observations, the
detection persists, either the eccentricity must be continuously pumped
by an outer planet or WASP12b's tidal dissipation constant must be significantly
larger than the typically adopted value ($10^6$).  Since, the presence of an additional
planet should be detectable via long term radial velocity monitoring, these
two scenarios are distinguishable and thus this system might allow 
detailed constraints to be placed on $Q_P$, an important, but difficult to measure, parameter.  
Additionally, if an second planet is causing a non-zero eccentricity, 
the amount of internal tidal heating that must
be dissipated is of order $5\times10^{28}\frac{10^6}{Q_P}$~ergs~s$^{-1}$.  This is a large
amount of energy (2 orders of magnitude
greater than what is calculated by \citet{boden03} for HD~209458b)
which would have a significant effect on the radius of the planet.

WASP-12b is the hottest and largest transiting exoplanet yet detected.  It has
a mass, $M_{\rm pl}$=1.41~M$_J$, radius, $R_{\rm pl}$=1.79~R$_J$, and equilibrium planet
temperature, $T_{\rm eq}$=2516~K.  The planet orbits a bright late-F star with a temperature, $T_{\rm eff}$=6300~K
and radius, $R_{*}$=1.57~R$_{\odot}$.  The host star has significantly enhanced metallicity over
solar, is evolving off the zero age main sequence, and has an age of $\sim$~2~Gyr.  
Additional follow-up observations of this exciting
planet will address some of the most pressing questions in exoplanet
research, in particular what mechanism or mechanisms are causing the large observed
radii of some hot Jupiters and what effect does the stellar irradition and stellar metallicity
have on the atmospheric and structural properties of close-in gas giant planets.

\acknowledgements
The SuperWASP Consortium consists of astronomers primarily from the Queen's University Belfast,
St Andrews, Keele, Leicester, The Open University, Isaac Newton Group La Palma and
Instituto de  Astrof{\'i}sica de Canarias. The SuperWASP Cameras were constructed
and operated with funds made available from Consortium Universities and the UK's Science and
Technology Facilities Council. SOPHIE observations have been funded by the Optical Infrared
Coordination Network. The data from the Liverpool and NOT telescopes was obtained under the
auspices of the International Time of the Canary Islands. We extend our thanks to the staff
of the ING and OHP for their continued support of SuperWASP-N and SOPHIE instruments. 

The radial velocity observations have been funded by the
Optical Infrared Coordination network (OPTICON),
a major international collaboration supported by the
Research Infrastructures Programme of the
European Commissions Sixth Framework Programme.

This paper was based in part on observations made with the Italian Telescopio Nazionale 
Galileo (TNG) operated on the island of La Palma by the Fundación Galileo Galilei of the 
INAF (Istituto Nazionale di Astrofisica) at the Spanish Observatorio del Roque de los 
Muchachos of the Instituto de Astrofisica de Canarias

The Liverpool Telescope is operated on the island of La Palma by 
Liverpool John Moores University in the Spanish Observatorio del Roque de los 
Muchachos of the Instituto de Astrofisica de Canarias with financial support from the 
UK Science and Technology Facilities Council.

\label{lastpage}

\end{document}